\documentstyle[12pt,epsfig]{article}
\begin{document}
\def\beq{\begin{equation}}
\def\eeq{\end{equation}}
\def\beqa{\begin{eqnarray}}
\def\eeqa{\end{eqnarray}}

\title{On the CP asymmetries in Majorana neutrino decays}

\author{Esteban Roulet$^{1}$, Laura Covi$^{2}$ and Francesco
Vissani$^3$ \\  \  \\
 $^{1}${\it Depto. de F\'\i sica Te\'orica, Universidad de
Valencia},\\ {\it E-46100, Burjasot, Valencia, Spain} \\ $^{2}$
{\it School of Physics and Chemistry, Lancaster University}\\{\it 
Lancaster LA1 4YB, United Kingdom}
\\ $^{3}$
{\it Deutsches Elektronen Synchroton, DESY}\\{\it 
22603 Hamburg, Germany}}

\date{}
\maketitle

\begin{abstract}
We study the CP asymmetries in lepton number violating two body
scattering processes and show explicitly how they vanish, in agreement
with unitarity constraints. We relate these cross section asymmetries
to the CP decay rate asymmetries of the intermediate massive neutrinos
and show how the inclusion of the Universe expansion via Boltzmann
equations is the key ingredient to produce a non-vanishing asymmetry
in spite of the unitarity constraint on the cross sections. We then
show that the absorptive parts of both the one loop vertex and self
energy corrections  do contribute to the CP decay asymmetries.
\end{abstract}
\vfill
\thispagestyle{empty}

\newpage

To generate dynamically a cosmological baryon (or lepton) asymmetry,
two crucial ingredients are the existence of CP violation and the
required out of equilibrium condition (the third one being just the
existence of B or L violating interactions). We want in this note to
examine carefully these two issues with the aim of clarifying an
existing debate in the literature. We will concentrate in the
discussion of leptogenesis scenarios \cite{fu86}, in which a baryon asymmetry
results after reprocessing an initial lepton asymmetry, whose origin
is related to the existence of heavy Majorana neutrinos. However, our
conclusions are general and apply to all scenarios based on out of
equilibrium decays.

The CP asymmetries are usually computed by studying the decay rate
asymmetries of an unstable heavy particle. This asymmetry arises in
general through the interference of a tree level process with a one
loop diagram containing an absorptive part. One such diagram is
depicted in fig.\ 1a and gives the so-called `vertex' contribution. The
point in debate is whether the absorptive part of the `self-energy'
like diagrams (fig.\ 1b) also contributes to the partial decay
asymmetries. The existence of this kind of contributions was noticed
a few times \cite{ig79}-\cite{pi97}, but they are still not 
always included (see e.g.
refs. \cite{lu92}-\cite{pl97});
furthermore, 
it has been argued in a recent paper \cite{bu97} 
that the self energy contributions 
should cancel among themselves, 
and hence be irrelevant for the generation 
of a baryon (or lepton) asymmetry.

\begin{figure}[hb]
\centerline{\epsfig{file=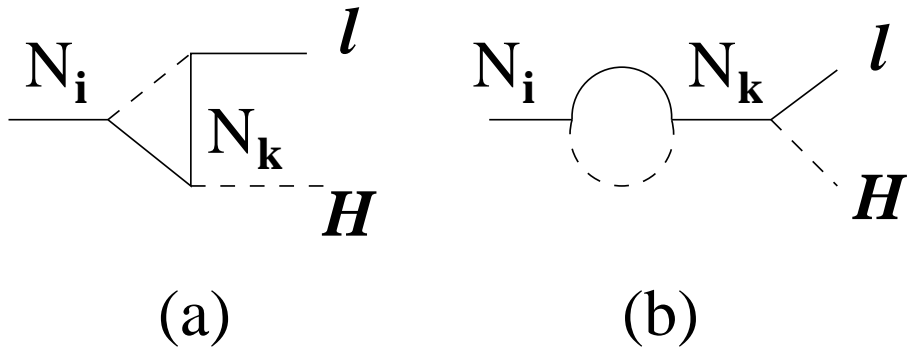,width=12cm}}
\caption{One loop diagram which interferes with the tree level
amplitude to give a CP asymmetry in the decay of a heavy singlet
neutrino $N$, coupled through a Yukawa interaction to the standard
lepton and Higgs doublets.}
\end{figure}

The usual argument invoked to include these absorptive self-energy parts
in the computation of the decay rates is that they cannot be reabsorbed
in the wave function renormalisation of the heavy states without
loosing the hermiticity of the Lagrangian, and hence they are physical. 
On the other side, 
in ref.~\cite{bu97} it was argued
that the heavy states, being unstable, 
are not a good starting point for 
perturbative calculations. 
In this spirit, it was suggested to use 
cross section asymmetries involving stable asymptotic states, in which
the heavy unstable particles appear only 
as virtual intermediate states. 
The following CP asymmetry was studied:
\beq
\epsilon_\sigma \propto\sigma(\ell^cH^*\to \ell H)-\sigma(\ell H\to
\ell^c H^*) .
\label{xsecasymm}
\eeq
(Here and in the following we will always assume that there is an
implicit sum in the flavour of the initial and final leptons, and in
the flavour of the intermediate heavy neutrinos.)
After computing in detail all 
the self-energy contributions to this
quantity, a complete cancellation was found. 
It was then concluded that only the vertex diagrams could provide a
non-vanishing contribution.
The results in~\cite{bu97} indeed reproduce what 
one would obtain just including the vertex
contributions to direct and inverse 
decay rate asymmetries. 

However, unitarity considerations  
imply that $\epsilon_\sigma$ defined in
eq.~(\ref{xsecasymm}) is actually exactly zero at the one-loop level.
In fact, unitarity implies that
 the probabilities for all possible transitions to and
from a state $i$ should sum to one, i.e.
\beq
\sum_j |M(i\to j)|^2=\sum_j|M(j\to i)|^2 ,
\label{unitarity}
\eeq
where the sum over $j$ includes all states and antistates. 
In the case
of leptogenesis, one may take $j=1,2,3$ to describe $\ell_j$ and
$j=4,5,6$ to describe $\ell^c_{j-3}$. Hence, 
 writing for brevity only the leptonic fields, the
asymmetry  $\epsilon_\sigma$  will be proportional to 
\beqa \epsilon_\sigma& \propto &
\sum_{i=4}^6\sum_{j=1}^3[\sigma(\ell_i\to \ell_j)-\sigma(\ell_j\to
\ell_i)]\cr
&=&
\sum_{i=1}^6\sum_{j=1}^3[\sigma(\ell_i\to \ell_j)-\sigma(\ell_j\to
\ell_i)]\cr
&=&
\sum_{i=1}^6\sum_{j=1}^3[\sigma(\ell_j\to \ell_i)-\sigma(\ell_j\to
\ell_i)]
 =0 ,
\label{zero}
\eeqa
where we first added and subtracted the lepton
conserving scattering processes and then used the
unitarity relation in eq.\ (\ref{unitarity}) appliied to the $2\to 2$
scatterings\footnote{Beyond $O(\lambda^6)$, where
$\lambda$ are the Yukawa couplings, final states with more than two
particles will appear in the unitarity relation. These multiparticle
states will also have to be included in the definition of
$\epsilon_\sigma$.}.

The non-zero result obtained in ref.~\cite{bu97} 
is due to the omission of some diagrams: those associated to a further 
tree level process, in which 
$N$ is exchanged in the $u$-channel\footnote{There 
are also one loop box diagrams, 
but they preserve lepton number
(as can be seen by hypercharge conservation), 
hence they do not contribute to the asymmetry 
considered. Loop diagrams in the 
$u$-channel exchange are irrelevant 
at leading order since they have no absorptive parts.}. 

There is a simple and instructive way to see pictorially how all the
different diagrams cancel out, which we first illustrate by
reproducing the cancellation of the self-energy parts found in
ref.~\cite{bu97}.
For this sake we will use the fact 
that the interference term in the cross section
(the only one contributing to the asymmetries) 
is proportional to ${\rm Re}[{\cal M}_0{\cal M}^*_1]$, 
where ${\cal M}_{0(1)}$ is the tree (one loop) amplitude.

Therefore, one can pictorially represent the self-energy contributions to
$\sigma(\ell^c H^*\to \ell H)$ as in figure~2, where the cut blobs
are the initial ($i$) and final ($f$) states, while the remaining blob
stands for the one-loop self-energy. This last is actually the sum of two
contributions, one with a lepton and one with an antilepton.

\begin{figure}
\centerline{\epsfig{file=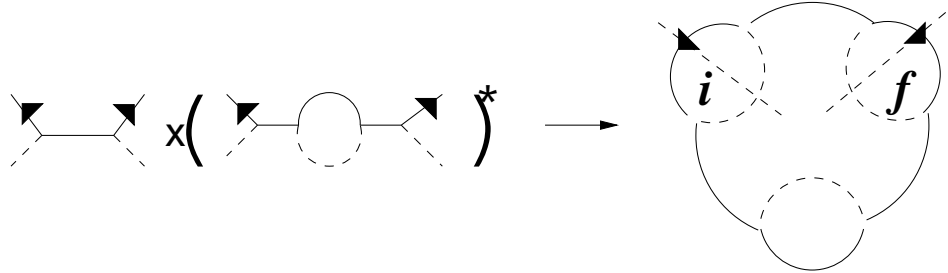,angle=270,width=12.5cm}}
\caption{Pictorial representation with cut blobs of the interference
term of the cross section $\sigma(\ell^c H^*\to \ell H)$. Higgs
scalars are represented with dashed lines. Initial and final leptons
with solid lines with arrows corresponding to the lepton number flow.}
\end{figure}

Now, in the computation of $\epsilon_\sigma$, only the absorptive part
of the loop will contribute, and the Cutkoski rule then tells us that the
self-energy blob should also be cut. Hence, putting explicitly
the two contributions to the self-energy and using the Cutkoski rule,
the asymmetry $\epsilon_\sigma$ will be proportional to the diagrams 
 in figure~3.
\begin{figure}
\centerline{\epsfig{file=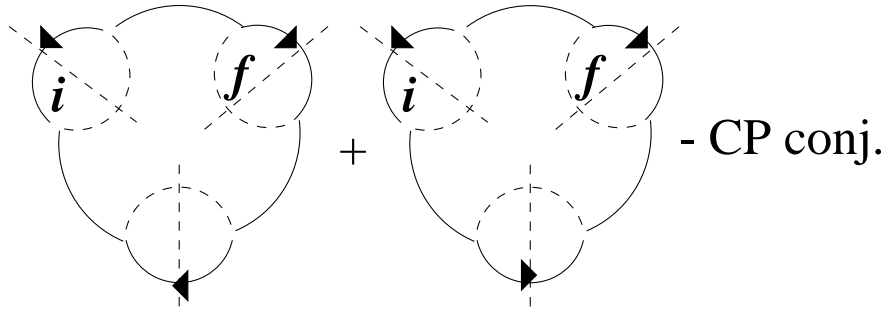,angle=270,width=12cm}}
\caption{Pictorial representation of the self-energy contribution to
the cross section asymmetry $\epsilon_\sigma$.}
\end{figure}

In the first diagram, the lepton is running clockwise in two blobs and
counterclockwise in the third blob, while in the second diagram only one
lepton is running clockwise while the other 
two go in the reverse direction (drawing 
conventionally the leptonic lines towards 
the external part of the diagram).
The CP-conjugate processes, which have to be subtracted to obtain the
asymmetry, have just the arrows reversed, and hence under conjugation
the two diagrams shown change into each other. This leads to
a complete cancellation of the asymmetry.

Turning now to the vertex contributions, to see the cancellations we
need to add the three contributions shown in figure~4 (including the
tree level $u$-channel interfering with the one-loop self-energy
diagram). 
\begin{figure}[bht]
\centerline{\epsfig{file=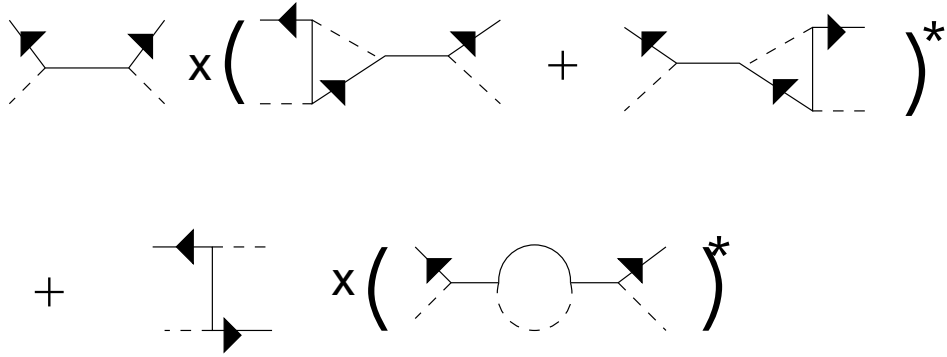,angle=270}}
\caption{Interference terms contributing to $\sigma(\ell^c H^*\to \ell
H)$ and involving the one loop vertex contributions as well as the
term necessary to enforce the cancellation of the CP asymmetry. }
\end{figure}
\begin{figure}[thb]
\centerline{\epsfig{file=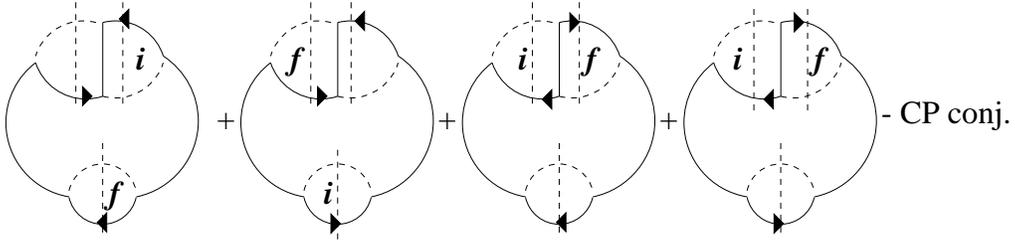,angle=270,width=13.6cm}}
\caption{Pictorial representation of contribution to
$\epsilon_\sigma$ from the diagrams in Figure~4.}
\end{figure}
In terms of cut diagrams, this can be expressed as in
figure~5, where CP-conjugate stands for the same four diagrams with
all the arrows reversed.  Hence, the CP-conjugate contribution will
exactly cancel the four diagrams, since changing the directions of the
arrows just exchanges among themselves the first and fourth diagrams,
as well as the second and third ones. We then see explicitly that the
absorptive part of the self-energies are also playing here a crucial
role, enforcing the cancellation of the CP violation produced by the 
vertex diagrams.

The only remaining diagrams to be considered are the interference of
the one-loop vertex diagrams with the tree level $u$-channel.
Pictorially, they are represented in figure~6, and they again cancel
since the two diagrams change into each other under CP-conjugation.
\begin{figure}[hb]
\centerline{\epsfig{file=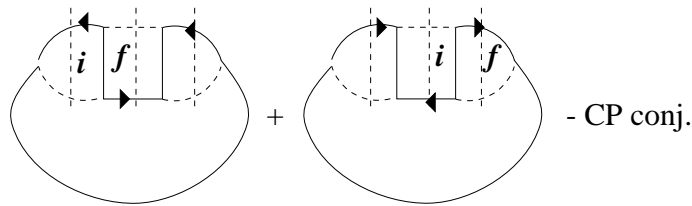,angle=270,width=9.3cm}}
\caption{Pictorial representation of the contribution to
 $\epsilon_\sigma$ from the interference of the $u$-channel tree
diagram with the one loop vertex corrections.}
\end{figure}

To further interpret these results, consider for simplicity the case of
two generations, and just the terms in the interference part of the
cross sections involving the exchange of two $N_1$ and one $N_2$.
First, for the interference contributions resulting 
from the self-energies diagrams depicted in figure~2, 
there will be three terms corresponding to:

$i)$ the interference of the $N_1$ tree exchange  with the  $N_1\times
N_2$ mixing.

$ii)$ the interference of the $N_1$ tree exchange  with the  $N_2\times
N_1$ mixing.

$iii)$ the interference of the  $N_2$ tree exchange  with the  $N_1\times
N_1$ mixing.

If we consider the resonant production, i.e. with $s\simeq M_1^2$, the
contribution in $i$) will factorise into the production of a real $N_1$
and its decay through mixing with $N_2$ (see ref.~\cite{bu97}).
Similarly, $ii)$ will correspond to the production of a real
$N_1$ through mixing with $N_2$ and its subsequent decay. However,
$iii$) does not factorise, since $N_2$ is off-shell, but as shown above
its presence is crucial to cancel the CP asymmetry in the process.

The same classification 
applies to the contributions in
figure~4, and there will 
be real $N_1$ production with CP violating
decay; CP violating inverse 
decay producing a real $N_1$ which then
decays (these first two contributions involving the vertex
corrections); and then a 
non-factorisable term coming from the $N_2$
$u$-channel exchange interfering 
with a self-energy diagram with two
$N_1$s. 

{}From the previous discussion it is clear 
that no contribution to the cross
section CP asymmetry (\ref{xsecasymm})
should survive, neither from the 
diagrams involving self-energies nor from those 
involving vertices. 
These vanishing results do not however forbid
the possibility of obtaining  
a net cosmological lepton asymmetry:
The crucial point is to take into account 
the expansion of the Universe, following, 
by the Boltzmann equation, the evolution 
of the heavy neutrinos and 
of the lepton number distributions.
In this way, the different processes just discussed ($i$ to $iii$),
will take place at different times, and the expansion of the Universe
will affect the external conditions at production and decay 
so as to make the cancellation
among the different pieces no longer complete (see below). Here is where the
departure from equilibrium becomes essential, since in equilibrium no
asymmetry can be produced just due to the unitarity constraint
 \cite{we79,ko80}. 

Let us consider in slightly more detail
the evolution of the lepton number density.
The off-shell contribution from  
CP violating $2\to 2$ scatterings 
($iii$ above) can be included in the Boltzmann equation 
exploiting the constraint from eq.\ (\ref{zero}),
that is subtracting from 
the total cross section asymmetry
(which is zero) the asymmetry associated with real 
intermediate states ($i$ and $ii$ above)
\cite{offshell}. 
This implies that the asymmetry produced 
in conditions of thermal equilibrium 
by inverse decays ($ii$) and 
scatterings processes $(iii$) 
is cancelled by the asymmetry produced by 
direct decays\footnote{In the case of 
large hierarchies in the spectrum of the
heavy states, the only particle relevant for the 
generation of the lepton asymmetry
is typically the lightest singlet neutrino $N_1.$
In the general case, however, an interplay 
between all processes takes place, and
the sign of the lepton number asymmetry generated 
by any of the individual process $i,ii$ and $iii$  
depends on the heavy state considered.}.
But the first asymmetry can be erased, 
at least partially, by L violating 
scatterings (not necessarily CP violating) 
{\em before} the actual decays of the states $N_1$ take place. 
It is important here that the real intermediate states 
 actually live for a long time by cosmological standards. As a
consequence,  
between the production vertex and the decay one the temperature
changes sufficiently as to allow first the compensation of the initial
L asymmetry by L violating scatterings, which then by the time of the decay
had already gone out of equilibrium. In this way, a non-vanishing asymmetry
can eventually be generated.

This scenario requires a 
CP violating asymmetry in 
the decay of the $N_1,$ 
which at lowest order 
is provided by the interference of 
the tree level diagrams with 
one loop diagrams that have non-zero absorptive parts.
Hence, in a Boltzmann equation treatment of the $N_i$
distributions, both contributions of 
the vertex {\em and} of the self-energy 
diagrams should be included. 

\bigskip

It is a pleasure to thank F.\ Botella, E.A.\ Paschos,
N.\ Rius, A.\ Santamaria and A.Yu.\ Smirnov 
for useful discussions. 
This work was partially supported by CICYT, Spain,
under grant No.\ AEN-96/1718.

\vfill\eject

\end{document}